# Initial growth behavior and resulting microstructural properties of heteroepitaxial ZnO thin films on sapphire (0001) substrates

C. Liu,[a)] S. H. Chang, and T. W. Noh[b)]

*ReCOE & FPRD, School of Physics and Astronomy, Seoul National University,*

*Seoul 151-747, Korea*

M. Abouzaid and P. Ruterana

*SIFCOM UMR 6176 CNRS-ENSICAEN, 6 Boulevard du Marechal Juin, 14050 Caen*

*Cedex, France*

H. H. Lee

*Pohang Accelerator Laboratory, POSTECH, Pohang, 790-784, Korea*

D.-W. Kim

*Department of Applied Physics, Hanyang University, Ansan, Kyunggi-Do 426-791,*

*Korea*

J.-S. Chung

*Department of Physics and CAMDRC, Soongsil University, Seoul, 156-743, Korea*

---

[a] E-mail: cliu@phya.snu.ac.kr
[b] E-mail: twnoh@snu.ac.kr

We have investigated the initial growth behavior and resulting microstructural properties of heteroepitaxial ZnO thin films prepared by pulsed laser deposition on sapphire (0001) substrates. High-resolution x-ray diffraction and transmission electron microscopy studies revealed that the initial growth behavior and the microstructure of the films significantly depend on the growth parameters. ZnO films grown at 700 $^o$C with 20 mTorr $O_2$ partial pressure initiated with a columnar growth mode and contained two kinds of domains. These domains were in-plane orientated either ZnO[$11\bar{2}0$]//Al$_2$O$_3$[$10\bar{1}0$] or ZnO[$10\bar{1}0$]//Al$_2$O$_3$[$10\bar{1}0$], and were surrounded by highly defective domain boundaries with threading dislocations. The films grown at 800 $^o$C with 1 mTorr $O_2$ showed 2-dimensional layered growth with only one in-plane epitaxial relationship, ZnO[$11\bar{2}0$]//Al$_2$O$_3$[$10\bar{1}0$]. Most of the defects in the layered grown films were basal plane stacking faults near the interface between ZnO and the substrate. The formation mechanism of the 30$^o$-twisted domains with the in-plane orientation of ZnO[$10\bar{1}0$]//Al$_2$O$_3$[$10\bar{1}0$] is discussed.

Due to its superior excitonic properties, ZnO has attracted enormous research attention focusing on its applications toward the short wavelength light emitting diode and exciton-based laser diode.[1,2] ZnO thin films have mostly been grown on sapphire ($Al_2O_3$) (0001) substrate due to its hexagonal symmetry and low cost, but the very large mismatch of lattice constants and thermal expansion coefficients between ZnO and sapphire makes it difficult to control and to understand the initial growth behavior. The crystalline quality and device properties are affected by the lattice strain relaxation and accompanying structural evolution in the initial growth stage.[3,4,5] Despite many x-ray diffraction (XRD) studies,[6,7,8,9,10] the initial growth behavior and the microstructures of heteroepitaxial ZnO thin films on sapphire are not very well understood. The ZnO films grown by radio-frequency (RF) sputtering were proposed to follow the two dimension (2D)-three dimension (3D) transition mode in which 3D columnar grains nucleated on the highly strained and well aligned 2D epitaxial ZnO initial layer formed on the substrate.[6,7] Conversely, the ZnO films deposited by molecular beam epitaxy (MBE) were reported to nucleate into 2D islands on the substrate, and the films contained a high crystalline epitaxial layer on top of a highly defective thin interface layer.[8,9] This apparent inconsistency implies that the initial growth mode and resulting structural properties can be greatly affected by the growth methods/conditions. Therefore it is important to systematically investigate the initial growth behavior and the microstructures of the heteroepitaxial ZnO initial layers in order to clearly understand the growth mechanism and to control the initial growth for high quality ZnO epilayers.

In this letter, we report a detailed study on the initial growth behavior and the microstructural properties of ultrathin ZnO films deposited by pulsed laser deposition (PLD) on sapphire (0001) substrates, using high-resolution x-ray diffraction and transmission electron microscopy (TEM) measurements.

ZnO thin films were deposited on sapphire (0001) substrates by ablating a ceramic ZnO target (purity 99.999%). The sapphire substrates were annealed in a furnace at 1100 $^{\circ}$C for 1 hr in air ambient to obtain atomically flat surface. A KrF excimer laser with a wavelength of 248 nm was used, and its fluence and repetition rate were about 3 J/cm$^2$ and 3 Hz, respectively. The deposition temperature of ZnO was varied as 600, 700, and 800 $^{\circ}$C, and the O$_2$ partial pressure was 1, 10, and 20 mTorr. Two deposition conditions: (a) 700 $^{\circ}$C with an O$_2$ partial pressure of 20 mTorr and (b) 800 $^{\circ}$C with an O$_2$ partial pressure of 1 mTorr, were used to investigate the initial growth behaviors because the films grown at these two conditions showed typical (0002) ω-rocking curve profiles in the XRD measurements as reported.[6, 8] Films with thicknesses of 10, 20, and 40 nm were prepared at both 700 $^{\circ}$C / 20 mTorr and 800 $^{\circ}$C / 1 mTorr to investigate the initial structural evolution of heteroepitaxial ZnO films. The film thicknesses were estimated from x-ray reflectivity measurements. The high-resolution XRD was measured with a lab source (Bruker D8 Discover) and a synchrotron radiation source (Beamline 10C1, Pohang Light Source, Korea). TEM observations were made to investigate the microstructural properties of ZnO initial layers using a 002B Topcon high-resolution microscope operated at 200 kV.

Figure 1(a) shows (0002) ω-rocking curves of heteroepitaxial ZnO initial layers grown at 700 $^{o}$C/20 mTorr with the variation of film thickness. The 11 nm-thick ZnO film showed only a sharp peak with a full width at half maximum (FWHM) of about 0.02$^{o}$. When the film thickness increased to 18 nm, a weak and broad background peak appeared below the sharp peak. Further increase of film thickness to 41 nm resulted in a significant increase of the intensity of the broad background peak while the intensity of the central sharp peak increased slightly with no change in FWHM. Similar two-peak feature and thickness dependence of the ω-rocking curve were reported for heteroepitaxial ZnO thin films grown by radio-frequency magnetron sputtering. It was suggested that the central sharp component in the rocking curve was originated from two-dimensional (2D) ZnO initial layer which was well aligned to the sapphire substrate, whereas the broad component was from three-dimensional (3D) ZnO islands which were nucleated on the 2D initial layer. The origin of these two peaks will be further discussed later based on our detailed microstructural analyses. Figure 1(b) shows (0002) ω-rocking curves of ZnO initial layers grown at 800 $^{o}$C/1 mTorr with the variation of film thickness. As in the film grown at 700 $^{o}$C/20 mTorr, only a sharp peak was observed when the film thickness was 9 nm. However, the broad component in the rocking curve did not appear when the film thickness increased to 20 nm. Only weak Lorentzian tails appeared on both sides of the strong and sharp central peak when the film thickness was 39 nm. A similar rocking curve with weak Lorentzian tails was observed in a ZnO film grown by molecular beam epitaxy, and was assumed to be related to a two-layer structure within the ZnO film: a highly defective interface layer and a high crystalline epilayer on top of the defective interface. The apparently different evolution of the out-of-plane rocking curve with film thickness indicates that the initial growth mode and resulting crystallographic mosaicity of heteroepitaxial ZnO thin film can be significantly influenced by the growth parameters. It is noted that the broad components in the rocking curves in Fig. 1 became dominant with decreasing the growth temperature down to 600 $^o$C and with increasing $O_2$ partial pressure from 1 to 20 mTorr (not shown here). The lower growth temperature and higher $O_2$ partial pressure cause a reduction of adatom surface mobility/diffusion length during the film growth, leading to poor crystalline films.[11,12,13]

To investigate the in-plane lattice alignment, azimuthal phi (ϕ) scan measurements were performed for the ZnO films grown at different growth conditions, as shown in Fig. 2. Figure 2(a) shows the ϕ-scans along the azimuthal circles of ZnO {$10\bar{1}1$} reflection for the ZnO thin films grown at 700$^o$C/20 mTorr with the variation of thickness. It is noteworthy that there are two sets of peaks with an angle difference of 30$^o$ in the ϕ–scans. One set of peaks is originated from aligned heteroepitaxial ZnO domains with an in-plane epitaxial orientation of ZnO[$11\bar{2}0$]//Al$_2$O$_3$[$10\bar{1}0$],

which is typically observed in heteroepitaxial ZnO thin films on sapphire (0001) substrates due to the reduced lattice mismatch of 18 %. The other set of peaks is attributed to the formation of $30°$-twisted heteroepitaxial ZnO domains with an in-plane orientation relationship of ZnO$[10\bar{1}0]$//Al$_2$O$_3$$[10\bar{1}0]$, which is energetically unfavorable due to the large lattice mismatch of 32 %. The formation mechanism of the $30°$-twisted ZnO domains will be discussed later based on our TEM analysis. As the film thickness increases from 11 nm to 41 nm, the intensity of the peaks from $30°$-twisted ZnO domains relatively decreases and became very weak while the peaks from aligned domains becomes significantly strong and sharp. The relative evolution of the peaks in the ϕ-scans indicates that $30°$-twisted ZnO domains are metastable and can be dominant at the initial stage of film growth in our growth condition although aligned domains are energetically favorable in the heteroepitaxy and can overgrow the twisted domains with the increase of film thickness.

On the other hand, the ϕ-scans for the ZnO thin films grown at 800 °C /1 mTorr showed sharp peaks only from the aligned domains, as shown in Fig. 2 (b). The FWHM of the sharp aligned peaks was about $0.9°$, which is much smaller than that of the aligned peaks in the ϕ-scan of 41 nm-thick ZnO film grown at 700 °C/20 mTorr (about $2°$) [Fig. 2(a)]. This result indicates that heteroepitaxial ZnO films grown at higher temperature with lower O$_2$ partial pressure have better in-plane lattice alignment with only one in-plane orientation of ZnO$[11\bar{2}0]$//Al$_2$O$_3$$[10\bar{1}0]$.

The in-plane domain alignment at the early stage of film growth can affect the microstructure and strain status of the heteroepitaxial thin film because an in-plane twist between two domains on a substrate results in the formation of low-angle grain boundary which typically consists of edge-type threading dislocations.[14,15] The grain boundaries formed during the coalescence process of initial grains may develop inhomogeneous tensile strain in the film since the coalescence of grains reduces the surface energy at the expense of elastic tensile strain energy.[16] Thus, we measured the a- and c-lattice constants of the ZnO films by combining XRD θ-2θ scans of ZnO (0002) and K-scans of ZnO ($10\bar{1}2$) reflection to investigate the strain status of the films. Figures 3 (a) and 3 (b) show the a- and c-lattice constants of the ZnO films, respectively. Figure 3 (a) clearly indicates that the in-plane strain of ZnO films grown at 700 °C/20 mTorr was almost relaxed, while the films grown at 800 °C/1 mTorr were compressively in-plane strained, which is further supported by the consistent evolution of c-lattice constants with the film thickness in Fig. 3 (b). The heteroepitaxial ZnO thin films on sapphire substrates are expected to be compressively in-plane strained due to the large lattice mismatch and the thermal expansion mismatch. Since the ZnO films grown at 700 °C/20 mTorr showed poor in-plane lattice alignment with two different types of domains [Fig. 2(a)], we believe that the significant relaxation of compressive strain in the films can be mainly attributed to a compressive strain compensation by a tensile strain developed at low-angle grain boundaries with edge-type threading dislocations. The ZnO films grown at 800 °C/1 mTorr, which showed better in-plane alignment with only aligned domains [Fig. 2 (b)], might have less low-angle grain boundaries so that the compressive strain of the films is less compensated. It is noteworthy that the compressive strain of 11 nm-thick ZnO film grown at 700 °C/20 mTorr was almost relaxed while the 9 nm-thick ZnO film grown at 800 °C/1 mTorr was compressively strained, although both of them showed well-aligned out-of-plane mosaicity in the ZnO (0002) ω–scans (Fig. 1). This implies that the in-plane lattice alignment is more correlated to the strain relaxation mechanism than the out-of-plane alignment of heteroepitaxial ZnO initial layers.

To further understand initial growth behaviors and the microstructures of the ZnO thin films, high-resolution transmission electron microscopy (HRTEM) measurements were performed. The HRTEM image in Fig. 4(a) shows that the ZnO film grown at 700 °C/20 mTorr has a columnar structure formed by two kinds of domains. The dark areas, marked "1", correspond to the aligned domains, where the ZnO layer was imaged along its [$11\bar{2}0$] zone axis. The ABAB… stacking of the hexagonal close-packing along the [0001] direction can be seen in the magnified image, shown in Fig. 4(c). The bright areas, marked "2", correspond to the 30°-twisted domains, where the ZnO layer was imaged along the [$10\bar{1}0$] zone axis, and only the 0002 fringes were observed from these domains, as shown in Fig. 4(c). The selected area diffraction pattern, presented in Fig. 4(b), also shows two sets of domains with different in-plane epitaxial orientations, which is consistent with the XRD result [Fig. 2(a)]. It is noteworthy that both aligned and 30°-twisted domains grow directly on the substrate [Fig. 4(a)], implying that the formation mechanism of the twisted domains can be related to the adsorption site preference of adatoms on the substrate surface during the first monolayer deposition.

In contrast, TEM studies of ZnO film deposited at 800 °C/1 mTorr showed only one in-plane epitaxial orientation relationship of ZnO[$11\bar{2}0$]//Al$_2$O$_3$[$10\bar{1}0$], which is also in good agreement with the XRD result [Fig. 2(b)]. Figure 4(d) shows the HRTEM picture of ZnO initial layer deposited at 800 °C/1 mTorr, indicating that most defects are stacking faults lying in the (0001) basal plane near the interface between ZnO and the substrate,[17] some of which are marked with arrows. Away from the interface, the TEM picture shows well-layered atomic stacking. The 2D layered growth mode at 800 °C and 1 mTorr O$_2$ appears to have enough adatom mobility and migration time to form the thermodynamically more stable state without forming the 30°-twisted domains at the initial stage of film growth.

The formation of 30°-twisted domains in the initial ZnO thin films might be explained as a competition between the local interface energy and epitaxial strain energy.[18,19] The fact that the 30°-twisted domains are formed at lower temperature, although they are heteroepitaxially unfavorable due to the large lattice mismatch, implies that on the atomically flat and clean substrate the interface energy of the adsorption sites for the 30°-twisted ZnO domains are lower than that for the aligned ones. At the very beginning of the deposition, an isolated adatom will preferentially adsorb at the energetically lowest site on the sapphire surface. However, as the adatom coverage increases to form an epitaxial monolayer, the adsorption sites for aligned domains will become energetically more favorable because of the smaller lattice mismatch-induced epitaxial strain energy. Therefore, the adatoms will have a driving force to shift toward the adsorption sites for aligned ZnO domains. When the growth temperature of ZnO decreases, the kinetic process of the adatoms' diffusion to the adsorption sites for aligned ZnO domains will be suppressed, resulting in the formation of $30^{\circ}$-twisted domains at the initial stage of growth.

The formation mechanism of $30^{\circ}$-twisted domains proposed above assumes the sapphire surface is atomically flat and clean. We also deposited ZnO thin films on as-received sapphire substrates without furnace annealing. The ZnO thin films on as-received sapphire substrates did not show $30^{\circ}$-twisted domain peaks in the $\phi$–scan spectra over the investigated growth temperature range. The as-received sapphire surfaces might not have localized and strongly directed dangling bonds that favor strong adatom-substrate chemical binding, which would result in lower diffusion energy barriers of adatoms due to weaker interactions between adsorbate and substrate. Therefore, the adatoms on the as-received sapphire substrates might easily diffuse from the initial adsorption sites to epitaxial strain energy favorable adsorption sites to form aligned ZnO domains.

The microstructural properties revealed by TEM study also provide reasonable explanation for the evolution features of the $\omega$-rocking curves (Fig. 1) and for the different strain states (Fig. 3).

Although the out-of-plane rocking curve and its evolution with film thickness [Fig. 1(a)] show apparently similar features with those from the RF sputtered films, the columnar structure directly formed on the substrate surface [Fig. 4(a)] indicates that our PLD-grown ZnO did not follow the 2D-3D transition mode in which 3D columnar grains nucleate on highly strained and well aligned 2D epitaxial layers. Our XRD results showed that the sharp single peaks in the out-of-plane rocking curves (Fig. 1) do not always indicate the formation of atomically well-stacked high quality 2D epitaxial single layer. The occurrence of the broad components in the ω-scan profiles (Fig. 1) might be related to the lateral overgrowth and coalescence processes of thermodynamically more favorable columnar grains. As for the different strain states in Fig. 3, the significant relaxation of compressive strain in the columnar structured films are mainly attributed to the defective columnar domain boundaries with high density of threading dislocations as well as misfit dislocations at the interface between the substrate and ZnO film. It is noted that the domains in the films grown at 700 $^o$C/20 mTorr have relatively smaller in-plane domain size and two different types of in-plane epitaxial orientations with poor in-plane domain alignments, compared to those in the films grown at 800 $^o$C/1 mTorr with enhanced 2D layered growth mode and less domain boundaries.

In summary, we have investigated the initial growth behavior and the microstructures of ZnO thin films deposited on sapphire (0001) by pulsed laser deposition. The deposition of ZnO films at relatively lower temperature with higher $O_2$ partial pressure (700$^o$C/20 mtorr $O_2$) formed a columnar domain structure consisting of two types of domains with in-plane epitaxial orientations of ZnO[$11\bar{2}0$]//Al$_2$O$_3$[$10\bar{1}0$] and ZnO[$10\bar{1}0$]//Al$_2$O$_3$[$10\bar{1}0$]. The 30º-twisted domains formed at the early stage of film growth and were metastable while the aligned domains were thermodynamically stable. The deposition of ZnO films at relatively higher temperature with lower O$_2$ partial pressure (800ºC/1 mtorr O$_2$) favored 2D layered growth mode, resulting in highly crystalline strained epitaxial layers with only one in-plane epitaxial orientation.

This study was financially supported by the Korean Science and Engineering Foundation (KOSEF) through the Creative Research Initiative (CRI) program. The experiments at the Pohang Light Source were supported by the Ministry of Science and Technology (MOST) and the Pohang Steel Company (POSCO).

Figure caption:

FIG. 1. The (0002) ω-rocking curves of ZnO films grown at (a) 700 oC and 20 mTorr O2 and (b) 800 oC and 1 mTorr O2.

FIG. 2. The φ-scans for the {$10\bar{1}1$} reflection of ZnO films grown at (a) 700 oC and 20 mTorr O2 and (b) 800 oC and 1 mTorr O2.

FIG. 3. Thickness dependence of the (a) a-axis and (b) c-axis lattice constants. The dashed lines in (a) and (b) show the value of the a-axis and c-axis lattice constants in bulk ZnO, respectively. The solid line connecting the points is a visual guide.

FIG. 4. (a) High-resolution cross-sectional TEM picture of an 18 nm-thick ZnO film grown at 700oC/20 mTorr. The areas labeled "1" and "2" represent domains with ZnO[$11\bar{2}0$]//Al2O3 [$10\bar{1}0$] and ZnO[$10\bar{1}0$]//Al2O3[$10\bar{1}0$], respectively. The white arrows indicate the domain boundaries. (b) The selected area diffraction pattern from the corresponding crystallites. The spot lines labeled A and B belong to the ZnO[$11\bar{2}0$] and ZnO[$10\bar{1}0$] zone axes, respectively. The Al2O3 [$10\bar{1}0$] axis is labeled C. (c) Magnified TEM image of the rectangular area shown in (b). The ABAB…stacking is indicated in the image. (d) The HRTEM image of ZnO film grown at 800oC /1 mTorr. Two stacking faults have been underlined (arrows).

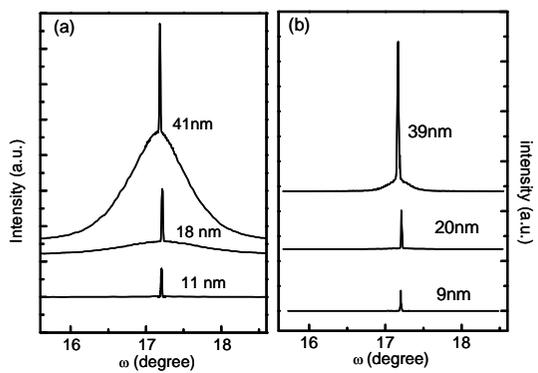

C. Liu *et al.*, Fig. 1

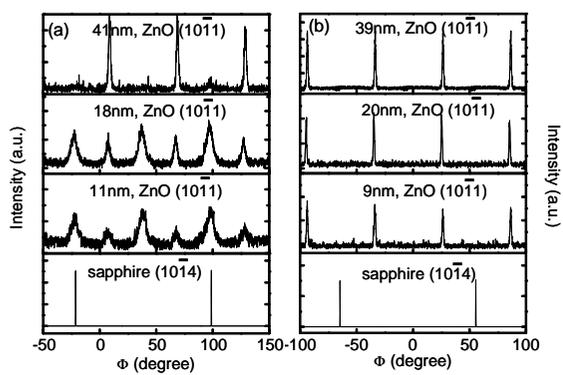

C. Liu *et al.*, Fig. 2

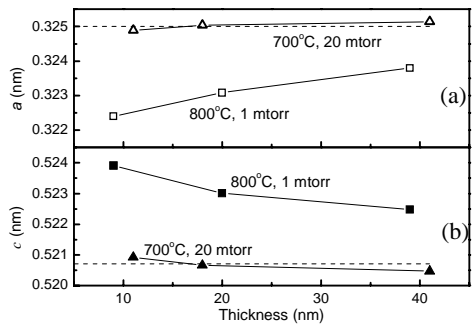

C. Liu *et al.*, Fig. 3

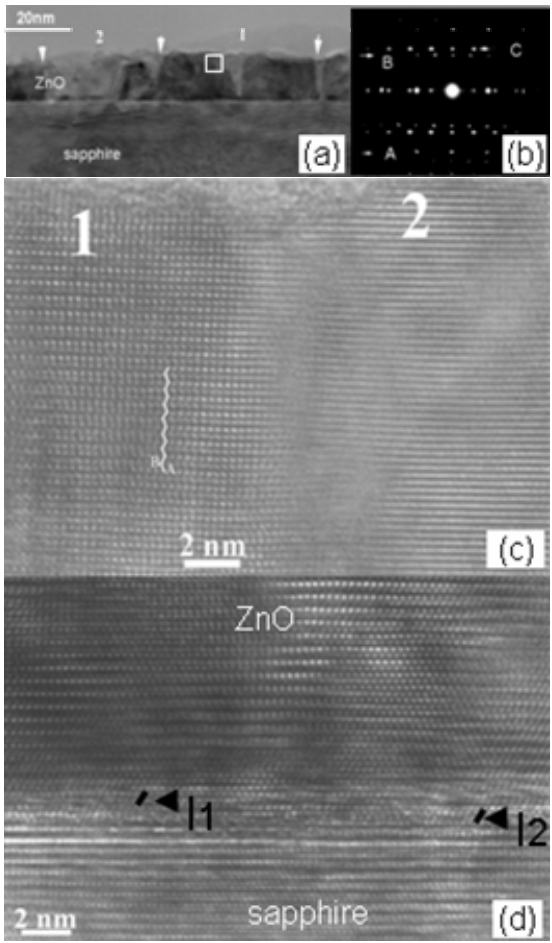

C. Liu *et al.*, Fig. 4